\documentclass[referee]{aa-mpa}
\usepackage{graphicx}
\makeatletter 
\let\@internalcite\cite
\def\cite{\@ifstar{\citeyear}{\citefull}}
\def\citefull{\def\astroncite##1##2{##1 ##2}\@internalcite}
\def\citeyear{\def\astroncite##1##2{##2}\@internalcite}
\def\citeau{\def\astroncite##1##2{##1}\@internalcite}
\def\citen{\def\astroncite##1##2{##1 (##2)}\@internalcite}
\def\possesivcite{\def\astroncite##1##2{##1's (##2)}\@internalcite}
\def\@citex[#1]#2{\if@filesw\immediate\write\@auxout{\string\citation{#2}}\fi
  \def\@citea{}\@cite{\@for\@citeb:=#2\do
    {\@citea\def\@citea{; }\@ifundefined
       {b@\@citeb}{{\bf ?}\@warning
       {Citation `\@citeb' on page \thepage \space undefined}}%
{\csname b@\@citeb\endcsname}}}{#1}}
\def\@cite#1#2{#1\if@tempswa , #2\fi}
\def\@biblabel#1{}
\makeatother

\begin{document}

\thesaurus{6(08.08.1, 08.12.1, 08.12.2, 08.16.3, 10.07.2)}

\title{Colour transformations for isochrones in the $VI$-plane}

\author{A.~Weiss\inst{1} \and M.~Salaris\inst{2}}

\institute{Max-Planck-Institut f\"ur Astrophysik,
           Karl-Schwarzschild-Str.~1, 85748 Garching,
           Federal Republic of Germany (aweiss@mpa-garching.mpg.de)
           \and
           Astrophysics Research Institute, Liverpool John Moores
           University, Twelve Quays House, Egerton Wharf, 
           Birkenhead L41 1LD, UK 
           (ms@staru1.livjm.ac.uk)
           }

\offprints{A.~Weiss}

\date{Received; accepted}

\authorrunning{Weiss \& Salaris:}
\titlerunning{Colour transformations for isochrones in the $VI$-plane}

\maketitle

\vspace{2.0cm}
\begin{abstract}
Reliable temperature--colour transformations are a necessary
ingredient for isochrones to be compared with observed
colour-magnitude-diagrams of globular clusters. We show that both
theoretical and empirical published transformations to $(V-I)$ to a large
extend exhibit significant differences between them. Based on these
comparisons we argue for particular transformations for dwarfs and
giants to be preferred. We then show that our selected combination of
transformations results in fits of $V$-$(V-I)$-CMDs of a quality
which is comparable to that of our earlier $V$-$(B-V)$ isochrones for
a wide range of cluster metallicities. The cluster parameters, such as
reddening, are consistent with those derived in $(B-V)$. Therefore, at
least in the case of the fit with our own isochrones -- based 
on the particular distance scale provided by our own horizontal branch
models, and on the treatment of convection by the mixing-length theory having
$l/H_p$ calibrated on our solar model -- the chosen transformations appear to
lead to self-consistent $(V-I)$ isochrones. Our isochrones are now well tested
and self-consistent for $B$, $V$ and $I$ photometric data.
\end{abstract}
\vspace{1.0cm}

\centerline{\it accepted for publication in Astronomy \& Astrophysics}
\clearpage

\section{Introduction}

The last step in comparing theoretical isochrones to globular cluster
colour-magnitude diagrams (CMD) is the transformation of luminosity
and effective temperature into brightness and colour. While there are a
number of both theoretical and empirical sets of transformations
available, this last step is, nonetheless, difficult and
critical. Known shortcomings in the theoretical transformations
which are obtained from theoretical non-grey model 
atmospheres calculations are, e.g.,
incomplete line lists for the opacity computation which affect the 
derived spectra and broad-band colours, the treatment of atmospheric 
convection which affects the colour determination as well (\cite{gcc:96}),
and an incomplete coverage of the $\log g$--$\log T_{\rm eff}$ and
composition parameter space, while empirical relations are naturally
based on a limited number of stars. 
The transformations of Kurucz and coworkers (\cite{kur:92};
\cite{cgk:97}) have well-known problems with the colours of red
giants (\cite{gcc:96}; \cite{sdw:97}; \cite{gva:98}), but are
nevertheless widely used. It is generally believed 
that the bolometric corrections, 
which are more important for most methods of cluster age determinations, are 
reliable, at least in their differential properties, and indeed results
based on different transformations are in reasonable agreement
(\cite{sdw:97}). 
As long as there are no unified stellar models available, which
treat the stellar interior and the non-grey atmosphere together in one
complete and realistic model (but see \cite{bern:98} for a first
unified solar model), such transformations are the only way to produce
observable photometric quantities.

There is also the persistent concern about the unsolved problem
of superadiabatic convection in the envelopes of cool stars, which
determines surface temperature and thus colour. In our own
calculations we have used the standard mixing-length theory with one
constant value for the mixing-length parameter, because even with this
very simplistic choice we are
able to match both the solar radius and the effective temperature of
metal-poor giants (\cite{sw:98}). The existence of a rather constant
value for this parameter is also supported by comparing stellar models
with the structure of convective envelopes computed by means of
2-dimensional hydrodynamical models (\cite{lfs:97};
\cite{frey:98}). However, there are other convection theories
possible (\cite{cm:91}), which lead to different $T_{\rm
eff}$ along an isochrone compared to the standard mixing-length
approach. The discussion in this paper is therefore restricted to our
own treatment of convection, and the conclusions could be different
for a different choice.

The confidence in ages derived from theoretical isochrones is higher
if the isochrone is able to reproduce the complete CMD from the lower
main sequence (MS) to the tip of the red giant branch (RGB) to high
accuracy, even if in practice only single points along the isochrone
may be used for determining the cluster age (e.g.\ turn-off and
horizontal branch).  Reliable colours become crucial for methods that
use colour differences, e.g.\ that between turn-off (TO) and RGB, for
(absolute) age determinations, for determining distance and age from
the MS-fitting technique, or for deriving correct integrated colours
from theoretical isochrones, which constitutes an necessary step for
studying old stellar populations in unresolved galaxies.

In our previous papers on this subject
(\cite{sdw:97}; \cite{sw:97}, ``SW97''; \cite{sw:98}, ``SW98'') we have 
taken great care to 
find transformations from the theoretical Hertzsprung-Russell-Diagram
(HRD) to the CMD in $V-(B-V)$, which give a satisfying fit for {\em
all} metallicities ranging now from ${\rm [Fe/H]} = -2.3$
to $-0.6$. We found that a combination of the transformations of
\citen{bk:78} and \citen{bk:92} (with an appropriate colour shift of 
the latter ones to enforce continuity in (B-V) at the common temperature
of 6000 K) satisfied our needs.

Work on globular cluster (GC) dating during the last few years has
received new attention. Due to improved models and methods
and lately due to HIPPARCOS-based distances to clusters, 
much lower ages are obtained than before, such that, at this time, no serious
conflict between cluster ages and the age of the universe exists
(SW98; \cite{chab:98}). 
In order to extend our work to a larger sample of
galactic halo clusters and in particular to those of the bulge (both needed
for questions about the formation of the Galaxy), it is necessary to
work in the $V-(V-I)$--CMD\footnote{If not stated differently, the
$I$-band refers to the Cousins system}, because there is now an 
increasing number of 
high-quality observations of clusters in $V-(V-I)$. 
However, up to now, well-tested isochrones for this colour have not 
been available.

In this paper, we demonstrate that existing empirical and theoretical
transformations provide very different $T_{\rm eff}-(V-I)$
relationships with only very few being in agreement with each other.
For the choice as to which transformation we apply to our isochrones, 
we want to see the following requirements being fulfilled:
(i) the transformation
should give RGB colours consistent with the determined age; (ii) an
isochrone that fits a cluster well in $(B-V)$, should do so in $(V-I)$,
too, for E(V-I) values compatible with the corresponding E(B-V).
In Sect.~2 we will present a number of recently published
theoretical and empirical transformations and apply them to our 
theoretical isochrones. We will show that considerable
differences in the resulting colours exist and that a priori 
there is no  preferred transformation to be selected. 
However, at the same time the 
reliability of the bolometric correction in $V$ will be reconfirmed.
We will provide arguments for our selection of transformations
for MS and RGB. In Sect.~3, we will then present
tests verifying its usefulness for the case of our own isochrones
applied to globular clusters of various metallicities.  By
no means this rather pragmatic approach is what one really  
wants for isochrone fittings, but at present it appears to be the only 
approach promising convincing results.  Finally, a discussion and 
summary closes the paper (Sect.~4).

\section{A comparison of different published colour-transformations}

\subsection{Transformation sources}

For the purpose of this investigation we have used a number of recently
published transformations, which we briefly describe in the following:

--\citeau{bk:78} (\citeyear{bk:78}; \citeyear{bk:92}, ``BK92''):
This is the source for synthetic colours we have
been using in our previous papers, showing that an appropriate
combination (henceforth called ``BK'') yields satisfactory CMD-fits
for all metallicities in the $V-(B-V)$ plane. 
The transformations are based on theoretical
atmospheres and are available for all metallicities and both for dwarfs
and giants. $(V-I)$ colours are available only for the
cooler stars on the lower MS and on the RGB (BK92).

--ATLAS9: This is a set of spectra and colours obtained from a model 
atmosphere grid computed with the ATLAS9 code (\cite{kur:92}),
covering the range of O-K stars. The first set of
ATLAS9 colours was discussed in \citen{kur:92}. The convective models
of this set were later revised (``K95''; see \cite{bcp:98}); the influence of the
treatment of convection in the ATLAS9 model atmospheres on some colour
indices has been discussed by \citen{cgk:97}.
The latest grid of ATLAS9 colours, computed from
``no-overshoot'' models is presented in \citen{bcp:98} for the
solar metallicity. In this paper we will use this latest set of colours
(hereinafter ``BCP98'') extended to lower metallicities by Castelli 
(1997, private communication). 

--\citeau{lcb:97} (\citeyear{lcb:97}, \citeyear{lcb:98}; ``LCB''): 
Synthetic spectra and colours based on various sets of 
theoretical atmospheres, among them K95 (covering almost completely the range of
gravities and  $T_{\rm eff}$ spanned by our isochrones). The theoretical
spectra for the solar metallicity have been corrected to
match empirical colour-$T_{\rm eff}$ relations; since comprehensive
empirical calibration data have only been available for the full
temperature sequences of solar-abund\-ance giant and dwarf stars, 
the spectra (and derived colours) of lower metallicity models have
been consequently corrected in such a way as to preserve the
monochromatic flux ratios predicted by the models.

-- \citeau{hab:99} (\citeyear{hab:99}; ``Next-Gen''): Grid of
theoretical model atmospheres, colours and bolometric corrections
covering the main sequence of globular clusters, with $T_{\rm eff}$
ranging between 3000 and 10000 K, $\log g$ between 3.5 and 5.5 and [M/H]
between -4.0 and 0.0. The atomic line list is the same as in
\citen{kur:92}, but the treatment of molecular, bound-free and
free-free opacity sources, as well as the equation of state and the
selected solar metal distribution are different. For convection,
standard mixing-length theory with a constant parameter of 1.0 was
used.

Note that all theoretical work up to now is for solar metal ratios
only, i.e.\ no enhancement of the $\alpha$-elements has been
considered. Work along this line is in progress (Castelli 1999, in
preparation). 

--\citeau{aam:96} (\citeyear{aam:96}; ``AAM96''): Empirical determination of the
$T_{\rm eff}$-colours relation for main sequence
stars as a function of metallicity, using a large sample of dwarfs
and subdwarfs. Since AAM96 use the Johnson--$I$-band, we have
transformed Johnson--$(V-I)$  into Johnson--Cousin--$(V-I)$ colours
according to \citen{fer:83} throughout this paper. 

--\citeau{mfo:98} (\citeyear{mfo:98}; ``M98''): Empirical determination of bolometric
correction-colour, colour-colour and $(V-K)-T_{\rm eff}$
relations obtained from photometric
observations of 6500 GC giants covering metallicities from ${\rm
[Fe/H]} = -2.2$ to $0.0$. The results are averaged over
two mean metallicity ranges below (``poor'') and above (``rich'') ${\rm 
[Fe/H]}= -1.0$. 

--\citeau{bcm:98} (\citeyear{bcm:98}; ``BCM98''): Empirical colour--$(V-K)$ relations for
giants in GC of low and intermediate metallicity as
well as one open cluster. These relations, together with an empirical 
$T_{\rm eff}-(V-K)$ relation taken from the literature,
provide direct $T_{\rm eff}$-colour transformations for GC
giants, taking into account their dependence on metallicity. 

In the empirical work, $\alpha$-element enhancement is considered only
as far as the observed stars show it. There is no systematic
differentiation between solar-type and $\alpha$-enhanced populations.

\subsection{Isochrones in the 
$V-(B-V)$--CMD and the bolometric correction}

The theoretical isochrones used throughout this paper were
computed for metallicities ranging from ${\rm [Fe/H]}=-2.3$ up 
to ${\rm [Fe/H]}=-0.6$, and are 
based on the same stellar evolutionary tracks as in our previous papers
(e.g.\ SW98), except for the fact that meanwhile we have 
extended all the calculations to the tip of the RGB, i.e.\ to
the onset of the core helium flash for all masses considered. The ZAHB
models have been calculated from RGB-tip models.
All isochrones\footnote{The isochrones are available on request from
the authors.} used here are for $\alpha$-element enhanced mixtures
($\langle{\rm [\alpha/Fe]}\rangle$=0.4).

\begin{figure}
\centerline{\includegraphics[draft=false,scale=0.60]{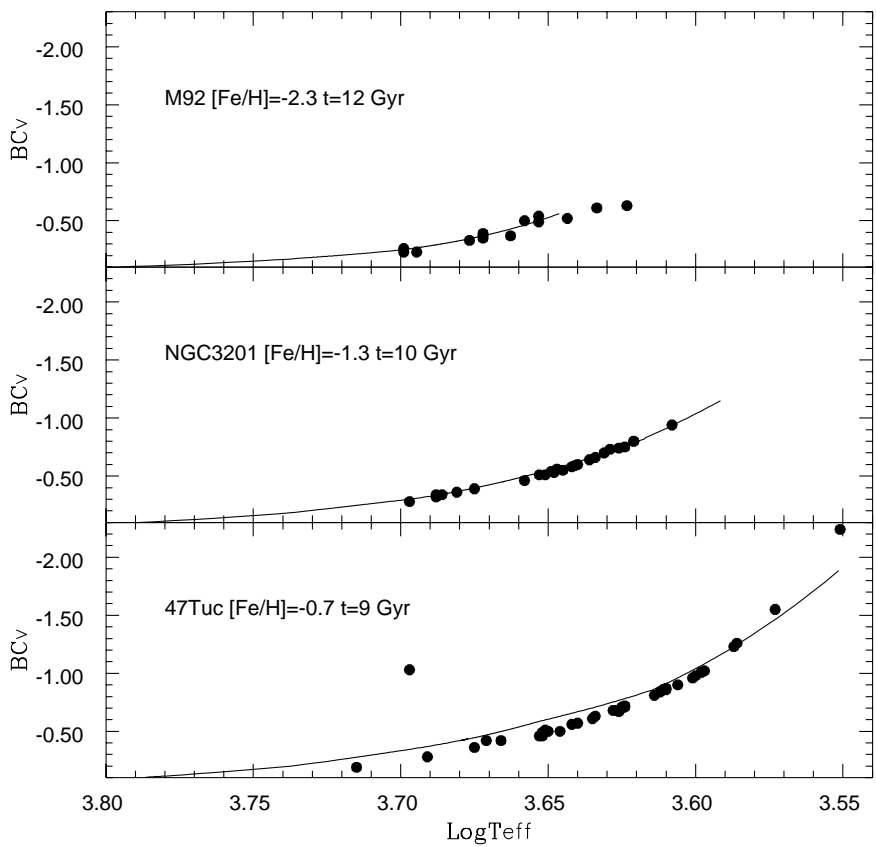}}
\caption[]{Comparison of $BC_{V}$ from BK with empirical data by 
\citen{fpc:81} (see text). The outlier in panel c is in the original
data}
\protect\label{figBC}
\end{figure}

As previously mentioned, our source for $(B-V)$ colours and bolometric
corrections ($BC_{V}$) is BK. In the previous papers (e.g.\ SW98)
we have shown that by using these colours we were able to obtain a
good fit to all clusters sequences covered by our isochrones,
to derive GC reddenings in good agreement with the results from 
\citen{z:85}, and to reproduce the HIPPARCOS subdwarf colours given
by \citen{gfc:97}.
As for the $BC_{V}$ scale, the zero point was calibrated to the
empirical scale by \citeau{aam:95} (\citeyear{aam:95}, \citeyear{aam:96}) for 
metal-poor MS stars; it reproduces the empirical
$BC_{V}$-values for the whole range of metallicities and MS 
temperatures covered by the isochrones within $\pm$0.02 mag.  
We required in particular that the solar $V$ magnitude
($V_\odot$ = 4.82) is reproduced.

With all the isochrones extended now up to the RGB-tip, we can also
compare our $BC_{V}$ scale for RGB stars with the empirical results by
\citen{fpc:81}. In Fig.~\ref{figBC} (panels a-c) we show the
comparison between the empirical $BC_{V}$ values derived for three
template clusters (M92, NGC3201, and 47~Tuc) spanning almost the whole
metallicity range of galactic GC and the $BC_{V}$ scale from our
theoretical isochrones. After correcting for the slightly different
(by 0.04 mag) zero points (i.e.\ to identical solar bolometric
correction), we find very good agreement between the two sets of
$BC_{V}$.  Also the $BC_{V}$ scales derived from the ATLAS9, Next-Gen 
(for the main sequence) and
LCB transformations agree well with the BK one. In particular,
once the $BC_{V}$ zero point is calibrated as we did for BK (see above
and SW98), the different scales agree within 0.05 mag all along the
isochrones and on the ZAHB; moreover, the brightness differences
between TO and ZAHB for fixed age and metallicity agree again within
0.05 mag. This confirms once more the consistency of the absolute ages
derived in SW97 and SW98.

\begin{figure}
\centerline{\includegraphics[draft=false,scale=0.50]{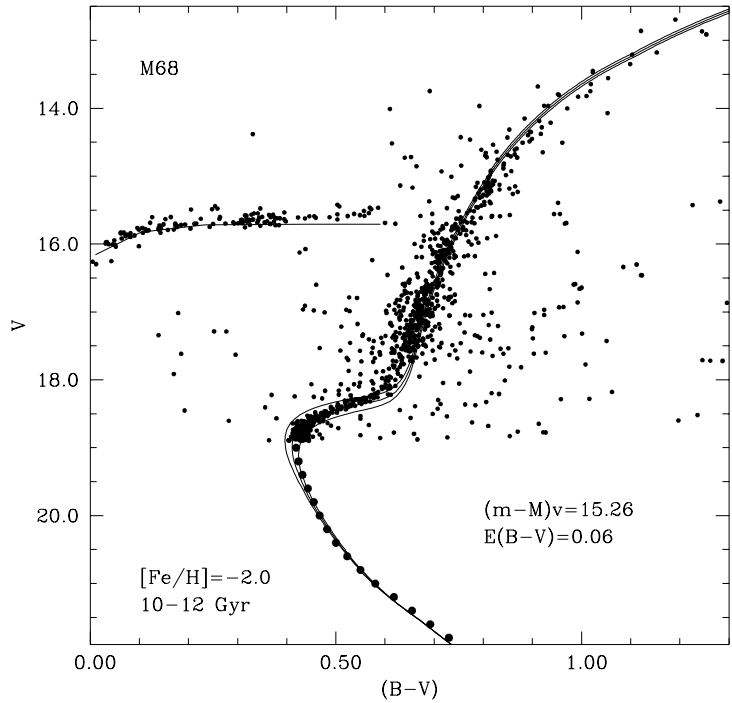}}
\caption[]{The $V-(B-V)$--CMD of M68 (data from \cite{wal:94}) and
isochrones of 10--12 Gyr.
For stars fainter than 19th mag only the
ridge line is shown; for stars brighter than that only the data points}
\protect\label{figM68bv}
\end{figure}

\begin{figure}
\centerline{\includegraphics[draft=false,scale=0.50]{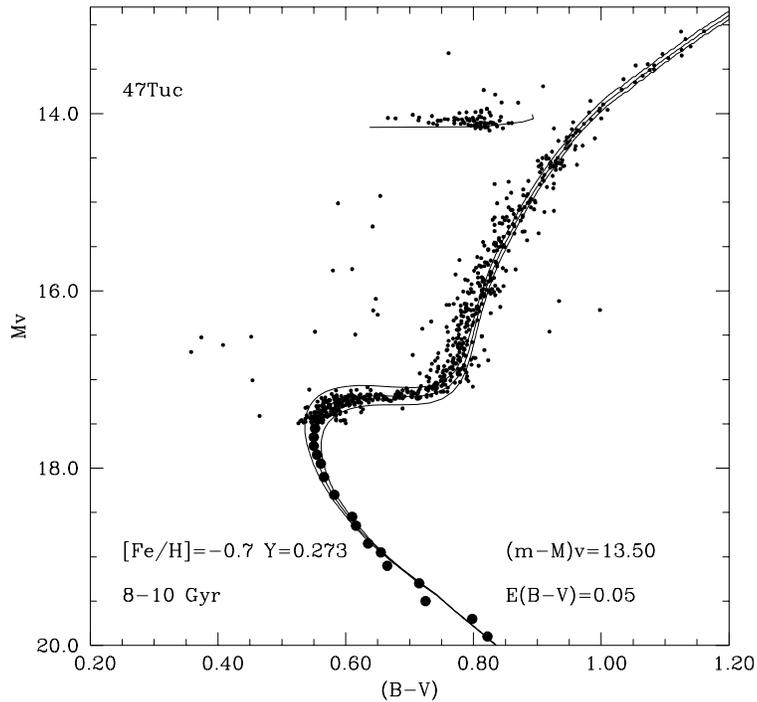}}
\caption[]{The $V$-$(B-V)$-CMD of 47~Tuc (data from \cite{kws:98})
with isochrones of 8--10 Gyr. For MS stars only the ridge line is shown.}
\protect\label{fig47Tucbv}
\end{figure}

As a final demonstration of the usefulness of the BK-transformations
applied to our isochrones,
we show in Figs.~\ref{figM68bv} and \ref{fig47Tucbv} the fits to M68
and 47~Tuc in the $V$-$(B-V)$ plane with our extended isochrones (in
SW97 and SW98 only fits up to the HB-brightness were displayed). The data for M68
(\cite{wal:94}) are the same as in our previous papers, while we used
the latest photometry by \citen{kws:98} for 47~Tuc.  We briefly recall
that the ages of the clusters are derived from the brightness
difference between the ZAHB (defined as the lower envelope of the
observed HB stellar distribution; see SW97 for more details) and TO;
this permits to derive the cluster age independently of the knowledge
of the cluster distance and reddening (due to the horizontal nature of
the HB), but of course the theoretical ZAHB models provide a distance.
Therefore, the distance modulus is derived by comparing the
theoretical ZAHB level to the observed one, while the reddening is
obtained from fitting the unevolved theoretical MS to the cluster one.
For 47~Tuc we recover with the new photometric data exactly the same
age, reddening and distance modulus as in SW98, where we were using
the older composite photometry by \citen{hhv:87}. Note that we have
used a rather high helium content of $Y=0.273$ for 47~Tuc; this is, at
least partially, based on the requirement of consistent vertical and
horizontal (colour-dependent) age indicators (see SW98 for a full 
discussion on this matter). It is evident from 
the pictures that the extended isochrones reproduce well the observed
RGB of both clusters.

\subsection{Isochrones in the $V-(V-I)$ CMD}

For transforming our isochrones to the $V-(V-I)$ plane we have
tested all transformations mentioned above. The BK
transformations to $(V-I)$ colours do not extend to temperatures higher than
6000 K such that neither ZAHB nor TO models are
covered for all metallicities and ages. 
Moreover, it is not guaranteed automatically that if this set of 
models reproduces well the 
observations in one colour band, the same is true for any other colour.

\begin{figure}[ht]
\centerline{\includegraphics[draft=false,scale=0.60]{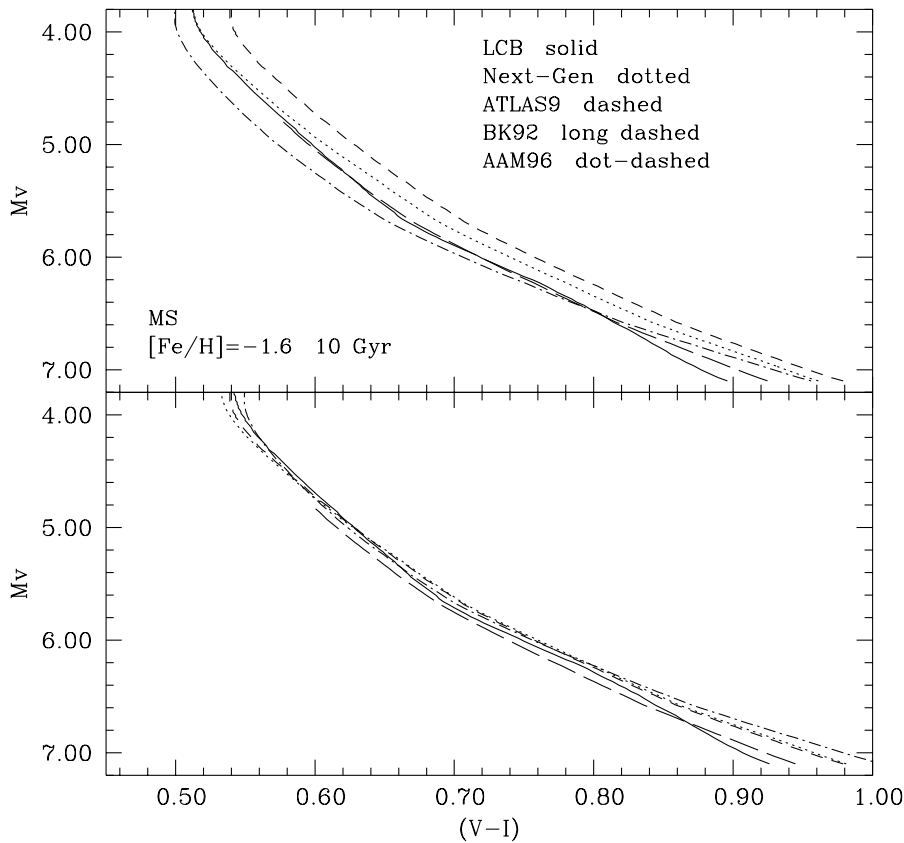}}
\caption[]{Upper panel: Result of various colour-transformations
(indicated in the figure) applied to a theoretical isochrone of ${\rm
[Fe/H]} = -1.6$ and 10~Gyr; shown is the main-sequence part. The
various labels are explained in Sect.~2.1.
Lower panel:
The same comparison, but now after adding a constant to
the various relations such that they reproduce $(V-I)_\odot=0.73$}
\protect\label{figMSvi}
\end{figure}

\begin{figure}
\centerline{\includegraphics[draft=false,scale=0.50]{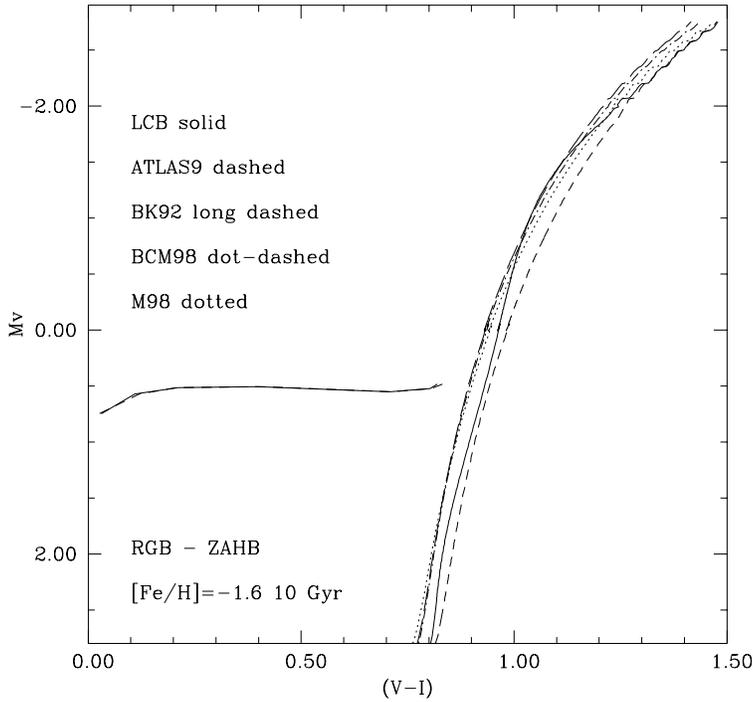}}
\caption[]{As the upper panel of Fig.~\ref{figMSvi}, but for the red
giant branch and ZAHB} 
\protect\label{figRGBvi}
\end{figure}

To exemplify the comparison between the different $T_{\rm eff}-(V-I)$ relations, 
we have used a representative isochrone 
(${\rm [Fe/H]} = -1.6$/10~Gyr), transformed to the $M_V$--$(V-I)$ plane in
Fig.~\ref{figMSvi} (upper panel) for the MS and in Fig.~\ref{figRGBvi} for the RGB. 
In the case of the BCM98 empirical transformations, we have used the 
empirical $T_{\rm eff}-(V-K)$ relation they give (\cite{rjw:80}) 
together with their empirically determined $(V-I)-(V-K)$ relations. As they
discuss in that paper, the \citen{rjw:80} relation is obtained
for solar metallicity stars, but there are good indications that it is not
affected by the metallicity.
As for M98, the authors derive an empirical $T_{\rm eff}-(V-K)$ relation for
metal-rich stars; following the arguments in BCM98, it appears to be
safe to use the same relation for the most metal-poor stars, too. In conjunction 
with the empirical $(V-I)-(V-K)$ relation we obtain again a direct 
$T_{\rm eff}-(V-I)$ transformation which is displayed in 
Fig.~\ref{figRGBvi}.

It immediately is evident from these illustrations that colour
differences up to between 0.05 and 0.10 mag exist all along the
isochrone and that there are no two transformations yielding the same
result for all evolutionary stages. This problem persists for other
metallicities as well and is, of course, not restricted to our
isochrones, but would be similar for any other.  Part of the mismatch
between the different transformations might be attributed to
differences in the composition (of model atmospheres or observed
stellar samples), to different or inappropriate assumptions in theory
(e.g.\ convection theory, line lists, LTE-conditions) and to the
internal spread in the empirical relations. Since the
width of the upper MS and the RGB in high-quality photometry is at
most as large as 0.05 mag, it is evident that whatever transformation
is chosen as it was published, isochrones can deviate from the CMD
structures by at least the width of the branches in colour. {\em It is
therefore an illusion to assume that a perfect isochrone fitting is a
strong indication for excellent theoretical stellar models or
isochrones!}

Recently, \citen{bcp:98} have independently compared a number of
theoretical colour transformations with empirical relations for MS
stars between spectral types O and M. Since they were interested in
aspects of population synthesis, they concluded that there is in
general an excellent agreement between theoretical and empirical
$T_{\rm eff}$--colour, bolometric correction--colour, and
colour--colour relations. However, inspecting their figures (e.g.\
Fig.~6, where the $T_{\rm eff}$--$(V-I)$ relation for the lower MS is
shown) one realizes that the deviations between the different sources
are of same order as in our case. The same spread is also visible in
the empirical $T_{\rm eff}$--$(V-I)$ relation (Fig.~2 as corrected in
their Erratum). Only due to the enormous scale of the colour axis the
agreement appears to be excellent. Globally, over all temperatures,
the agreement is indeed very good, but not sufficient for such narrow
structures as the RGB in galactic GC.

Along the MS, the empirical relation by \citen{aam:96} does not
support any theoretical transformation. The Next-Gen and ATLAS9
(BCP98) colours are the most similar, the average differences being
within 0.02 mag. Their $T_{\rm eff}-(B-V)$ relations, however, are almost
identical in the range of $T_{\rm eff}$ we are dealing with, and in
very good agreement with the $T_{\rm eff}-(B-V)$ relation adopted in SW97
and SW98, at least for $T_{\rm eff}$ larger than
$\approx 5300$~K. 

Assuming $(V-I)_\odot = 0.73$ as in \citen{bcp:98} we find that this value is
reproduced within 0.01 mag by BCP98, while Next-Gen, BK92, AAM96 and LCB yield
colours bluer by, respectively, $\approx$ 0.02, 0.02, 0.05, and 0.03 mag.
Within the uncertainty of solar colours (see, e.g. the discussions in  
\cite{chmi:81} and \cite{tayl:94}) all these values are probably
acceptable; nevertheless, for gaining more insight about the differences among the 
$T_{\rm eff}-(V-I)$ relations under discussion, we normalized the different colour 
transformations to the same value $(V-I)_\odot = 0.73$, by correcting
for the mentioned differences. After performing the necessary shifts
one obtains the lower panel of Fig.~\ref{figMSvi} (we have also tested
the case with ${\rm [Fe/H]} = -0.7$, with similar results). 
BCP98, Next-Gen (theoretical relations) and AAM96 (an empirical one) now
agree rather well with each other along the whole MS, while the 
other colours differ from these by different amounts at different
metallicities. This latter fact implies that the derivative $\delta
(V-I)$/$\delta [Fe/H]$ at a fixed value of the MS brightness, a
fundamental quantity needed for applying the MS--fitting method in the
$V$-$(V-I)$ plane, does depend on the particular colour transformation
used.  

From the results of these comparisons for the MS phase (which are 
independent of the underlying theoretical isochrones) it results that 
BCP98, Next-Gen and AAM96 $T_{\rm eff}-(V-I)$ relations reproduce acceptably well 
the solar constraint and, moreover, show the same differential behaviour.
If we base our 'objective' choice of the colour transformations  
on the consistency with empirical constraints and with independent
theoretical determinations, we have to conclude that 
BCP98, Next-Gen or AAM96 fulfill these criteria for the MS.

\begin{figure}[ht]
\centerline{\includegraphics[draft=false,scale=0.45]{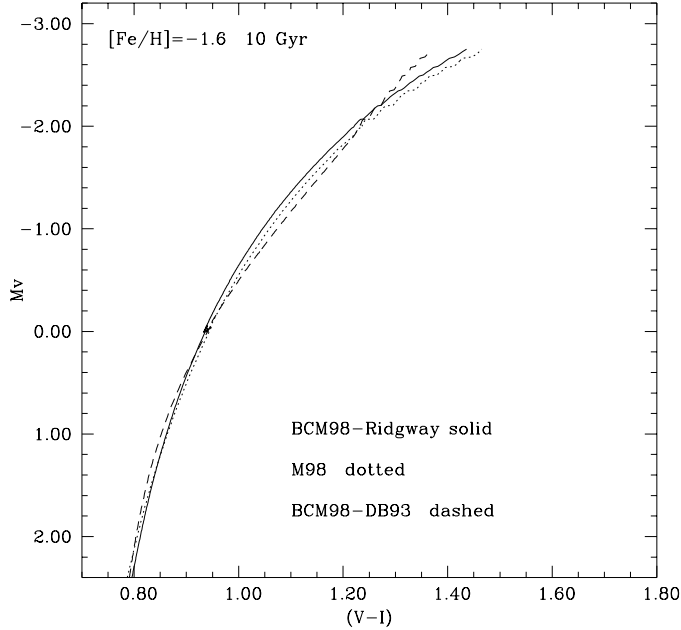}}
\caption[]{Comparison of RGB empirical $T_{\rm
eff}-(V-I)$ relations applied to our isochrones (see text)}
\protect\label{figRGBemp}
\end{figure}

While on the MS the isochrones transformed using different 
$T_{\rm eff}-(V-I)$ relations run more or 
less parallel to each other (except for LCB), along the RGB they 
differ also in slopes. However, we find that the two empirical relations 
tested on the RGB (M98 and BCM98) agree with each other rather well.
\citen{bcm:98} used the $T_{\rm
eff}$--$(V-K)$ relation by \citen{rjw:80} and their empirical $(V-I)$--$(V-K)$ relation.
These we combined to obtain the $T_{\rm eff}$--$(V-I)$ transformation. 
In Fig.~\ref{figRGBemp} we show how this relation (``BCM98-Ridgway'') 
compares to one using the more recent \citeau{ben:93}
(\citeyear{ben:93}; ``DB93'') $T_{\rm eff}$--$(V-K)$ relation instead and to the 
empirical one by \citen{mfo:98}. 
The agreement is, except for the brightest RGB part, better than 0.02 mag. 
The very good reciprocal consistency between these two
purely empirical and independently determined relations is appealing,
and on this ground we decided to try to use them for our isochrones.
In particular, we selected the BCM98-Ridgway relation 
which contains an explicit (albeit very weak) metallicity dependence.

\begin{figure}
\centerline{\includegraphics[draft=false,scale=0.45]{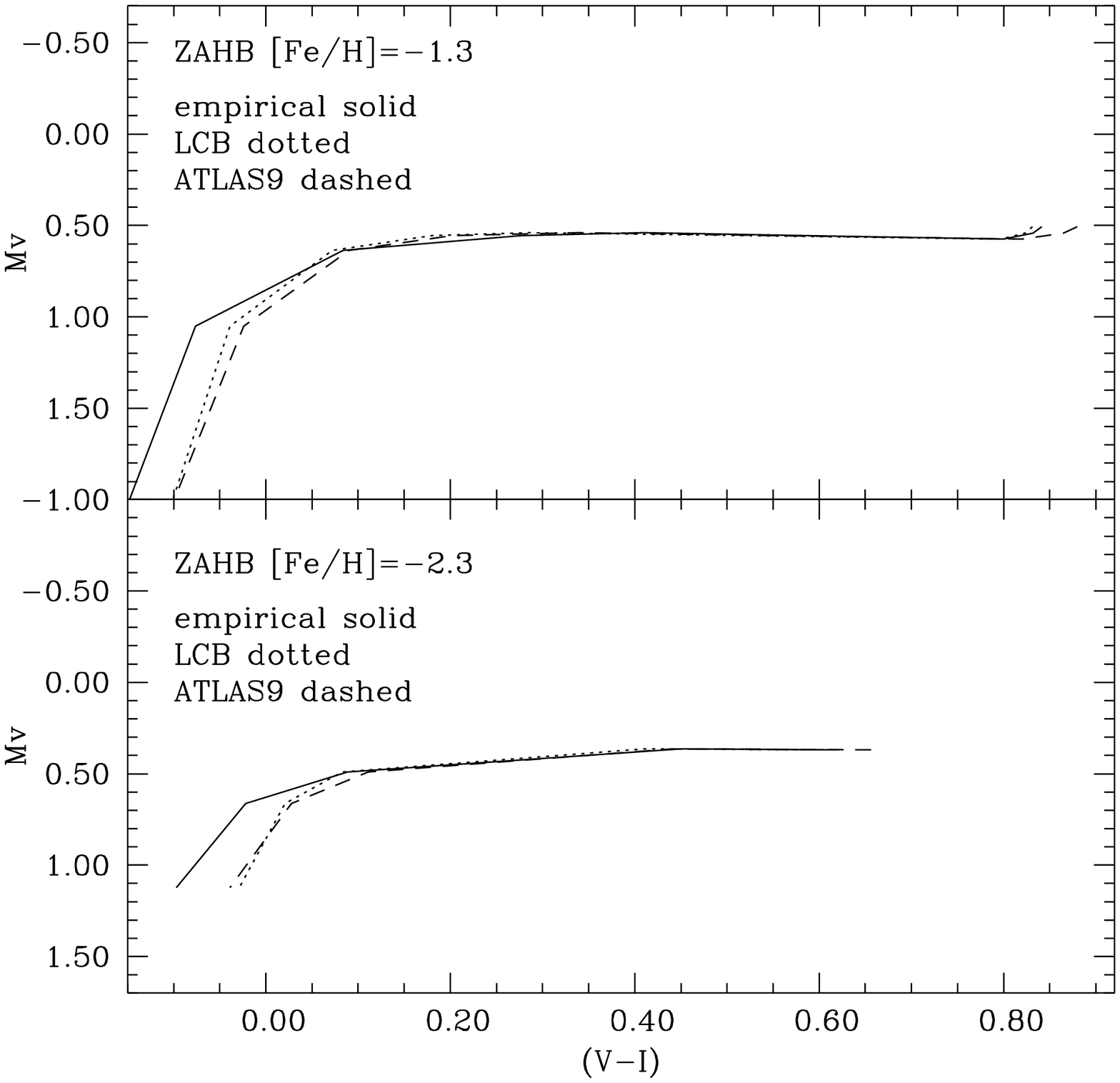}}
\caption[]{The ZAHB at two metallicities in $(V-I)$: shown is our
own relation (``empirical''; see text) and two other
transformations. Except for the most extreme colours the agreement is
good} 
\protect\label{figZAHB}
\end{figure}



We finally need a transformation for our HB models. We have
derived an empirical $(B-V)-(V-I)$ relation using 
multicolor photometries of clusters spanning the relevant range of
metallicities. We used HB data for M68 (\cite{wal:94}), M3
(\cite{fcc:97}), M5 (\cite{sbs:96}), and 47~Tuc (\cite{kws:98}), and we
applied this relation to our ZAHB models in the $(B-V)$ plane. In
Fig.~\ref{figZAHB} the comparison between the ZAHB transformed by
using this empirical colour-colour relation and two theoretical
transformations is shown for two different metallicities.  Evidently,
the theoretical relations agree quite well with each other in
predicting the turn-down of the ZAHB in the blue. But there are
differences not visible due to the horizontal nature of the HB. The
predicted colour along the HB differs in some cases by 0.05--0.10 mag,
just as for the RGB. Also, the ATLAS9 transformation yields
consistently redder stars at the cool edge (similar to the known
effect on the RGB) with increasing metallicity. The empirical relation
deviates at the blue end, possibly because of the effect that evolved
stars show up in the relation. Since observed HBs usually have a large
scatter at the blue, less luminous end, this part is not putting any
strong constraint on the ZAHB level, such that the difference between
the transformations as demonstrated in Fig.~\ref{figZAHB} is not
really significant. Except for the mentioned problem of too red
colours from the ATLAS9 transformation, we consider all relations as
being effectively equivalent. Since we have used empirical data from
those clusters we will discuss below, we will use the empirical
relation throughout the remainder of this paper.

\section{Globular cluster test cases}

In the last section we presented a possible choice of colour transformations
for all CMD-branches. This choice was made on the basis of which
relations agree best with each other. The real justification of this
choice, however, will be whether our transformation
successfully fulfills the
following requirements we are imposing: (i) for a given cluster, the
fits in $(B-V)$ and $(V-I)$ should be of the same quality; (ii) the
reddening obtained from both colours should be consistent; (iii)
derived quantities as age and distance should not depend on the colour
used. This implies that the final choice of transformations depends on
the set of isochrones used.  In the following tests we have employed
the BCP98 colours for the MS; between MS and RGB we smoothly switch
from the BCP98 $T_{\rm eff}-(V-I)$ relation to the BCM98-Ridgway one,
which overlap over a sufficiently wide range.  There is no need for
any additional correction or shift. For lowest metallicities they
almost agree (at the level of $\approx$ 0.01 mag).

To test our $T_{\rm eff}-(V-I)$-relation we use four GC of different
metallicity for which we have determined the age already in our
previous papers and for which $BVI$ data exist. We transform the
corresponding isochrones into $(V-I)$ by our new relation and
determine the reddening $E(V-I)$. For the transformation of reddenings
we take the extinction law by \citen{ccm:89}, which gives $E(V-I) =
1.3 \, E(B-V)$.

\begin{figure}
\centerline{\includegraphics[draft=false,scale=0.45]{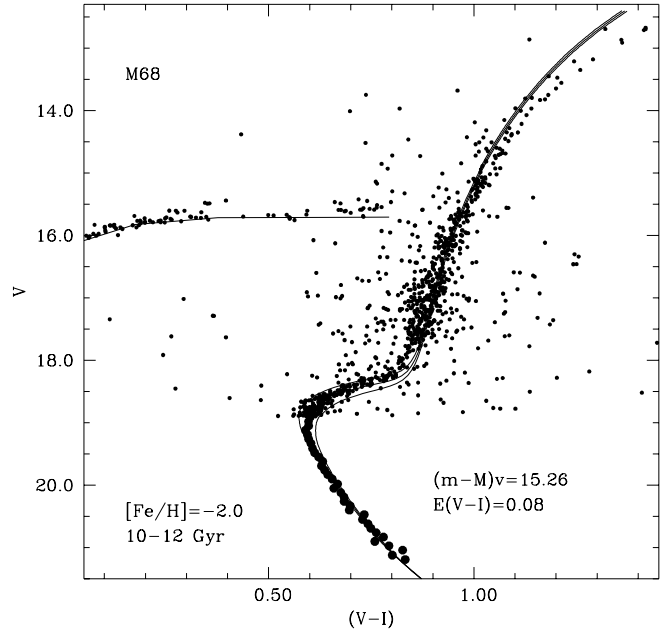}}
\caption[]{The $V-(V-I)$ isochrones for M68. Data are from
\citen{wal:94} and isochrone parameters as in \citen{sw:98}. For
simplicity, only data points brighter than $V=18.5$ are shown. Large
filled circles mark the ridge-line}
\protect\label{figM68vi}
\end{figure}

As an example for metal-poor clusters we use M68 (${\rm [Fe/H]} =
-1.99\pm0.10$; \cite{cg:97}); the $(V-I)$ data are from \citen{wal:94}. The
$(B-V)$-CMD has been shown in Fig.~\ref{figM68bv} and the corresponding
$(V-I)$-CMD is displayed in Fig.~\ref{figM68vi} (for sake of clarity
the ridge-line only is displayed for the MS).  The age determined
in SW98 from the $V$ brightness difference between TO and ZAHB
is $11.4\pm1.0$ Gyr, in good agreement with the 11 Gyr isochrone
displayed Fig.~\ref{figM68vi}. The MS ridge-line was determined by us
from the original data by taking the mode of the colour-distribution
within brightness bins.  The transformed isochrones appear to become
too blue in the brightest RGB parts, while the corresponding part in
$(B-V)$ is fitting very well\footnote{Worthey (1998, private
communication) points out that the \citen{wal:94} RGB-data are outliers
in $(V-I)$ vs.\ $(V-K)$ when compared to the standard giant branches
of \citen{dca:90} and the BCM98 data}. The reddening (obtained, as described
in \cite{sw:98}) is 0.06 mag in $(B-V)$ and 0.08 mag in $(V-I)$;
their ratio is as predicted by \citen{ccm:89}.

\begin{figure}
\centerline{\includegraphics[draft=false,scale=0.45]{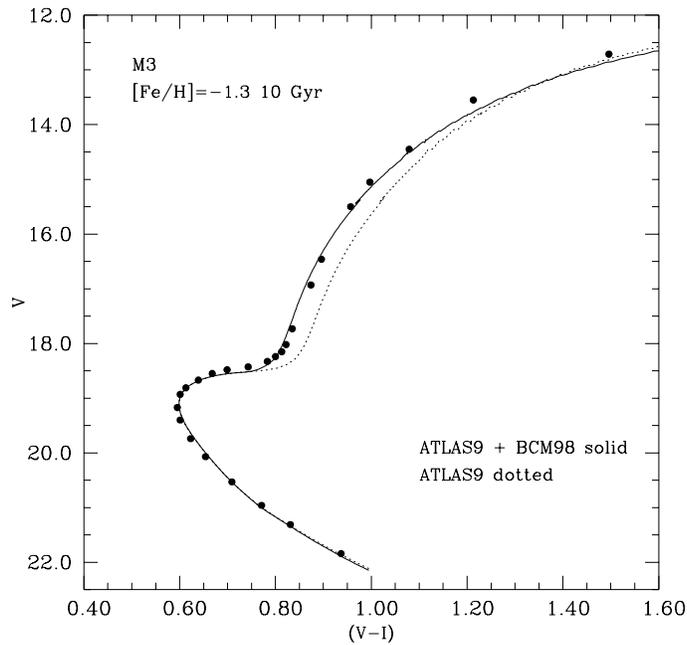}}
\caption[]{The $V$--$(V-I)$ 10-Gyr isochrone and the ridge line for
M3. Data are from \citen{jb:98}. Also displayed (dashed) is the same
isochrone transformed by using the ATLAS9 colours for the RGB as
well} 
\protect\label{figM3}
\end{figure}

As an example of an intermediate metal-poor cluster, Fig.~\ref{figM3}
shows the comparison between the ridge-line from \citen{jb:98} and the
10 Gyr-isochrone for M3 (${\rm [Fe/H]} = -1.34 \pm 0.06$;
\cite{cg:97}), for which SW98 determined differentially the age from
$(B-V)$-data (\cite{fcc:97}) to be $10.1\pm 1.1$ Gyr. The agreement is
excellent except for a small deviation in the upper part of the RGB.
The derived $E(V-I)$-reddening is 0.03 mag, in perfect agreement with
$E(B-V)=0.02$ (SW98; see also \cite{bcb:94}).  
Using the ATLAS9 $T_{\rm eff}-(V-I)$ relation for the
RGB as well results in a discrepancy of $\approx$0.05 mag for the
larger part of the upper part of the CMD. This difference, if taken
seriously, could be taken as evidence that the cluster should be older.

\begin{figure}
\centerline{\includegraphics[draft=false,scale=0.50]{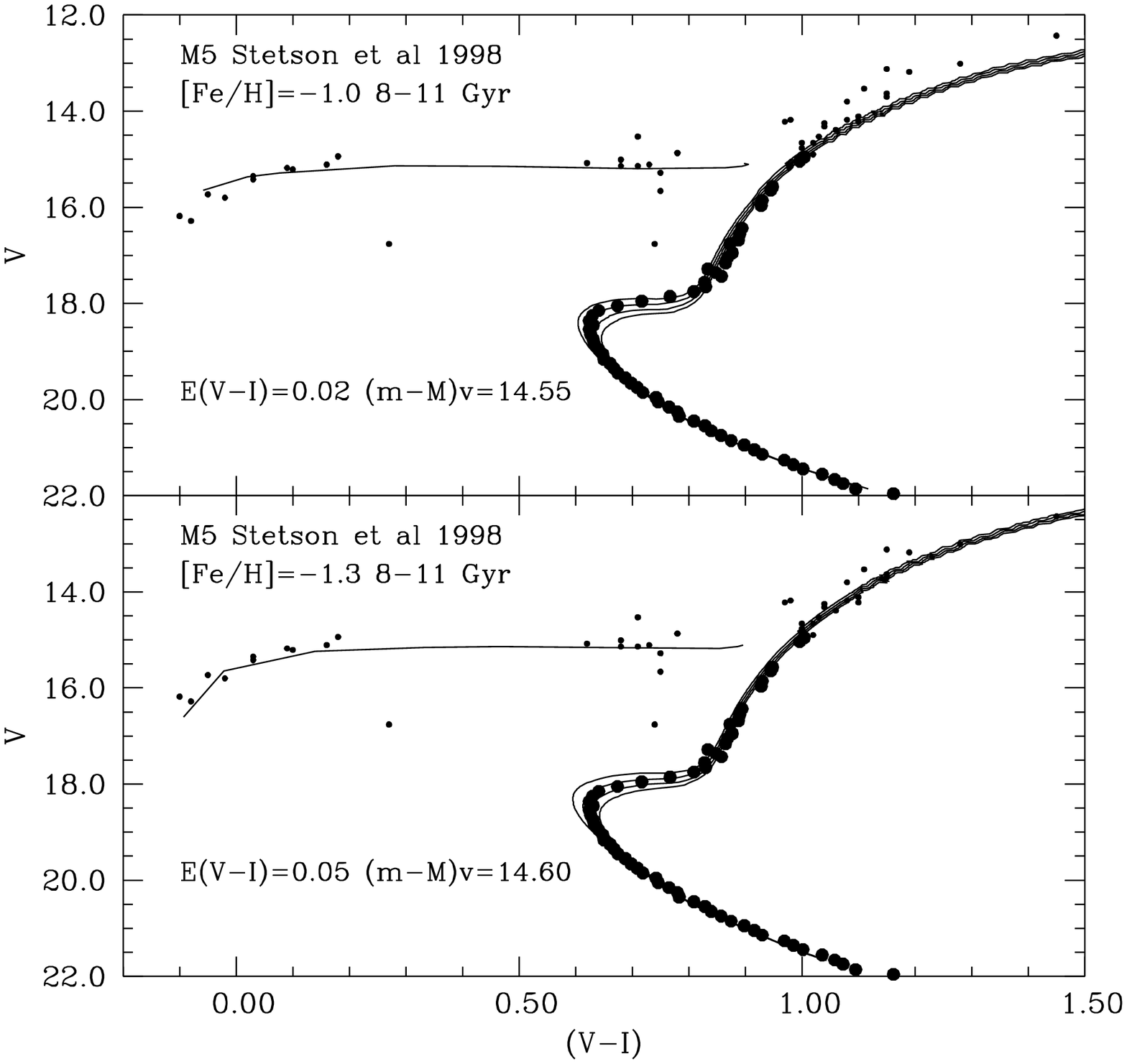}}
\caption[]{The $V$--$(V-I)$ ridge-line for M5 (from \cite{sbh:98}) and
isochrones of 8, 9 and 10 Gyr; upper panel: ${\rm [Fe/H]}=-1.0$;
lower: ${\rm [Fe/H]}=-1.3$}
\protect\label{figM5vi}
\end{figure}

From the third of our metallicity groups (see SW97) we have
selected M5; the absolute cluster age as derived from the $V-(B-V)$
diagram (\cite{sbs:96}) is $9.9\pm 0.7$ Gyr (SW98). Very recently
\citen{sbh:98} have published a new $V-(V-I)$ diagram of M5, and found
some discrepancy with respect to the scale and zero points of the
\citen{sbs:96} $V$ and $I$ magnitudes.  However, their estimated
TO-luminosity agrees with the one given by \citen{sbs:96} within 0.01
mag, and the same is true for the $V$ brightness of the HB and RGB
stars in common (with $V$ ranging between 14.5 and 16.5 mag).
Therefore, the absolute age as determined from the $V$ difference
between ZAHB and TO remains basically unchanged, and the same is true
for the cluster distance modulus determined from the fit of
theoretical ZAHB sequences to the observed one (in the following we
will therefore use the observed ZAHB-$V$ brightness as determined by
SW98 employing the more populated HB in the diagram by \cite{sbs:96}).
The \citen{sbh:98} $V-(V-I)$ diagram (they provide the ridge-line) is
shown in Fig.~\ref{figM5vi}. The two panels contain the comparison
with two of our isochrones, whose metallicities bracket the cluster
metallicity ${\rm [Fe/H]} = -1.11 \pm 0.11$, as determined by
\citen{cg:97}; the distance modulus is fixed by the observed ZAHB
brightness.  The quality of the fit is good in both cases; there is
only a difference in the upper part of the RGB, where the more
metal-rich isochrones are too red (as expected). By interpolating for
the value of the cluster metallicity we get $E(V-I)=0.03$, in good
agreement with $E(B-V)=0.02$ as determined by SW98 (due to a misprint
in Table~2 of SW98, the apparent cluster distance modulus was quoted
$(m-M_{V})=14.53$, while the correct value is $(m-M_{V})=14.57$).


\begin{figure}
\centerline{\includegraphics[draft=false,scale=0.50]{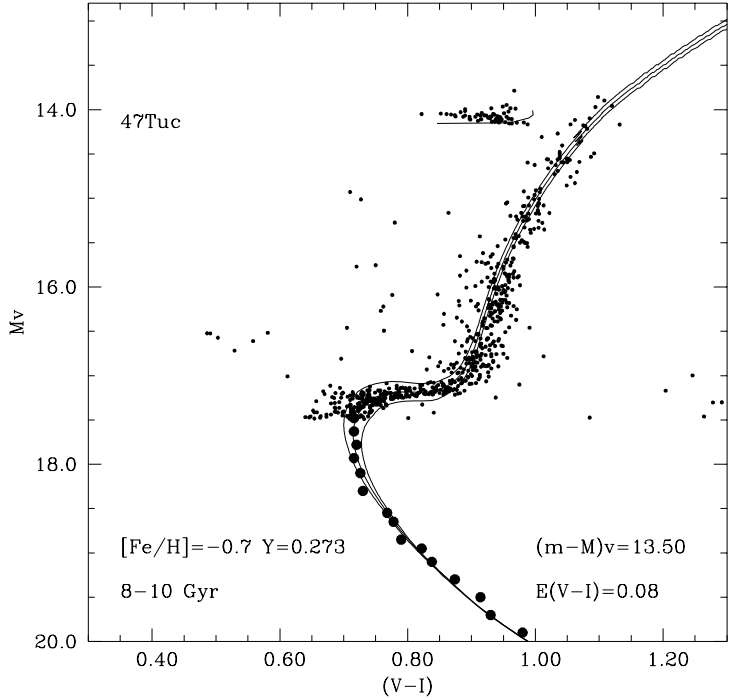}}
\caption[]{The $V-(V-I)$-CMD of 47~Tuc (data from \cite{kws:98})
with the same isochrones as in Fig.~\ref{fig47Tucbv}}
\protect\label{fig47Tucvi}
\end{figure}


Finally, we compare the data by \citen{kws:98} for 47~Tuc (${\rm
[Fe/H]} = -0.70\pm0.07$; \cite{cg:97}) with our isochrones. Since in SW98 we
used the data by \citen{hhv:87} for determining the age from the
$(B-V)$ data, we showed the same fit for the new data in
Fig.~\ref{fig47Tucbv} (the ridge-line shown for the cluster MS is
derived as previously described).  The theoretical isochrones are
identical to those in SW98 (including the higher helium content of
$Y=0.273$) except for the fact that they are now extended to the tip
of the RGB. The newly derived age, distance and reddening are the same
as in SW98. In Fig.~\ref{fig47Tucvi} the corresponding $(V-I)$-diagram
is shown. The reddening determined in this colour is 0.08, which, when
transformed to $E(B-V)$ is 0.06, differing from the expected value of
0.05 by only 0.01 mag. Note that while in $(V-I)$ the isochrone
appears to be a bit too blue for the lower RGB, it is too red by the
same amount in $(B-V)$. The fit is of the same quality in both colours
and slightly better than that shown in SW98 for the \citen{hhv:87}
data.

Inspecting the HB of 47~Tuc in $(V-I)$, one has the impression that is
inclined towards the blue. This effect is not so evident in $(B-V)$
(Fig.~\ref{fig47Tucbv}). We think it results from the fact that not
all stars have both $(B-V)$ and $(V-I)$ colours and that in $(V-I)$ a
number of redder HB stars are missing, thus making the inclination,
which we ascribe to evolution on the HB, more apparent. In any case,
the inclination could only be a brightness effect, independent of
colour. 

To conclude this section, we have demonstrated that the $T_{\rm
eff}$--$(V-I)$ transformation we have constructed on the basis of two
existing transformations, if applied to our own isochrones (SW98), 
results in cluster-CMD fits in $(V-I)$ which are equally good as in
$(B-V)$ for all metallicities and yield consistent reddenings in the two
colours. Thus our requirements formulated at the end of the
introduction and the beginning of this section are fulfilled.

\section{Conclusions}

In this paper we tried to approach the problem of fitting theoretical
isochrones to globular cluster data in $(V-I)$. This is important
since more and more $I$-band photometric data are becoming available and for
some objects (e.g.\ bulge clusters) will be the only data of quality
possible. Absolute ages of clusters determined by the turn-off
brightness or the $\triangle V$-method (see SW98) depend
almost exclusively on the bolometric correction $BC_{V}$. Theoretical
values agree, when properly calibrated, with empirical values and also
between different transformation sources.

On the other hand, transformations of effective temperatures to
colours differ from source to source by a rather constant level of
0.05--0.10 mag. From the comparisons we have shown it is evident that
(i) there is no straightforward way to decide which transformation is
the correct or best one and (ii) that the accuracy with which an
isochrone fits a colour-magnitude-diagram does not indicate the
quality of the underlying stellar evolution calculations, if the
mismatch is of the order of 0.05--0.1 mag. While in Sect.~2 we displayed
$(V-I)$--fits, the situation is the same in $(B-V)$ and corresponding
examples and conclusions can be found in the literature (e.g.\
\cite{gva:98}; \cite{dcm:97}).

The problem of inaccurate colours is not just a cosmetic one, but can
influence absolute age determinations which make use of colour differences
(e.g.\ that between turn-off and red giant branch), distances derived
by means of the main-sequence--fitting technique, or integrated
colours of unresolved stellar populations.  Finally, {\em if} a set of
transformation was available which is known to reproduce colours
accurately, any deviation of an isochrone from the observed CMD will
indicate a real inconsistency in the isochrone. As an example we
repeat that a colour mismatch of the RGB indicates (but does not
prove) a higher helium content for 47~Tuc (SW98). Thus, reliable colour
transformations are in principle a source for obtaining additional knowledge 
about cluster stars. We therefore attempted to find a preliminary solution
to this problem, such that we can continue our work about cluster
dating with $(V-I)$ data, expecting improved theoretical
transformations in the meantime.

We found that for MS stars the BCP98, Next-Gen and AAM96 
$T_{\rm eff}-(V-I)$ relations reproduce acceptably well 
the solar constraint and, moreover, show the same differential
behaviour with a rather constant offset of a few hundredths of a
magnitude. We selected the BCP98 colours for our models.

On the red giant branch, no theoretical transformation is
in agreement with empirical data over all the metallicity range spanned 
by our isochrones, but two recent empirical sources
(M98 and BCM98) confirm each other quite well. Since
BCM98 provide metallicity-dependent relations, we use this
one for the RGB. Both parts (MS and RGB) can be
connected smoothly along the subgiant branch.

For the ZAHB colours we have derived an
empirical $(B-V)-(V-I)$ relation using 
multicolor photometries of clusters spanning the relevant range of
metallicities, and we
applied this relation to our ZAHB models in the $(B-V)$ plane.

While the arguments just given have only led the way to our rather
pragmatic combination of two $T_{\rm eff}$--$(V-I)$ transformations,
the justification for the combined and final relation comes from the
tests we have performed using our own isochrones. For each of
the four metallicity ranges 
defined in SW97, which reach from the most metal-poor halo to
the half-solar metallicity disk clusters, we selected one cluster with
$BVI$-photometry (three of them had already been used for absolute age
determinations in SW98). We then showed successfully that (i)
the quality of the isochrone fit is comparable in both colours and
(ii) the reddenings determined independently from these two fits
fulfill the reddening law by \citen{ccm:89} either exactly or within
0.01 mag.

We therefore conclude that the combination of the (theoretical)
transformation of BCP98 for the main-sequence and the
(empirical) one by BCM98 for giants is of sufficient accuracy
as to allow us isochrone fits to globular cluster
colour-magnitude-diagrams in $V$--$(V-I)$.

In a forthcoming paper we will determine cluster ages based on
$VI$-photometry, both for clusters we already have investigated in our
previous papers and for clusters with $VI$-data only. Although $(V-I)$
is less sensitive to metallicity than $(B-V)$, we found that
$\triangle (V-I)$ (the colour difference between turn-off and red
giant branch, which we need for relative age determinations) at
fixed age is actually {\em more} metal-sensitive. 
Our approach of grouping clusters
into four metallicity bins will therefore not be accurate
enough. Either a finer metallicity grouping is needed (implying a
higher number of clusters for absolute age determination), or a purely
relative age approach is needed as in \citen{psw:98}. 

Independent of this, we are still
waiting for improved theoretical colour transformations, because for 
more metal-rich $\alpha$-enhanced clusters (e.g.\ bulge clusters) the BCM98 
results cannot be applied straightforwardly. The study of such clusters
therefore actually requires to some extent an extrapolation of our adopted colour 
transformations.

\begin{acknowledgements}
We are grateful to L.~Girardi and I.~Steele for helpful discussions
and E.~M\"uller for a careful reading of the manuscript. We
thank T.~Lejeune for providing an updated version of his
transformations in electronic form, and an anonymous referee for his
comments, which helped to improve this paper.
\end{acknowledgements}

\clearpage

\end{document}